\documentclass[prl,twocolumn,showpacs]{revtex4}
\begin{document}
\title{
   Pairing and Condensation
   of Laughlin Quasiparticles\\
   in Fractional Quantum Hall Systems}
\author{
   Arkadiusz W\'ojs,$^{1,2}$ 
   Kyung-Soo Yi,$^{2,3}$ and 
   John J. Quinn$^2$}
\affiliation{
   $^1$Wroclaw University of Technology, 50-370 Wroclaw, Poland\\
   $^2$University of Tennessee, Knoxville, Tennessee 37996, USA\\
   $^3$Pusan National University, Pusan 609-735, Korea}

\begin{abstract}
From the analysis of their interaction pseudopotentials, 
it is argued that (at certain filling factors) Laughlin 
quasiparticles can form pairs. 
It is further proposed that such pairs could have Laughlin 
correlations with one another and form condensed states
of a new type. 
The sequence of fractions corresponding to these states 
includes all new fractions observed recently in experiment
(e.g., $\nu=5/13$, 3/8, or 4/11). 
\end{abstract}
\maketitle

\section{Introduction}

In a recent experiment \cite{pan}, Pan {\sl et al.} observed the 
fractional quantum Hall (FQH) effect \cite{tsui,laughlin} at novel 
filling fractions $\nu$ of the lowest Landau level (LL).
Among other features, the FQH effect is a macroscopic manifestation 
of an incompressible character of the many-body ground state formed 
at a specific filling $\nu$. 
The new spin-polarized FQH states occur at filling factors outside 
the Jain sequence \cite{jain} of composite fermion (CF) states.
Some of them, as $\nu=4/11$ or 4/13, appear in the Haldane hierarchy 
\cite{haldane} of quasiparticle (QP) condensates, and therefore 
at first sight it might only be surprising that they have not been 
observed earlier.
However, the ``hierarchical'' interpretation of these states was 
questioned \cite{hierarchy} because of the specific form of the 
QP--QP interaction, which leaves their interpretation quite uncertain.
Others, such as the $\nu=3/8$ or 3/10 states, do not belong to 
the Haldane hierarchy, and thus the origin of their incompressibility 
is puzzling in an even more obvious way.
In the CF picture, these even-denominator fractions correspond to the 
half-filled first excited ($n=1$) CF LL.
While for the electrons half-filling their $n=1$ LL (including the 
double degeneracy of the lowest LL, this amounts to the total electron
filling of $\nu=2+1/2=5/2$) the FQH effect has been known for quite 
some time \cite{willet} (the incompressible $\nu=5/2$ state is known 
as a Moore-Read state \cite{moore}), the similar behavior of electrons 
and CF's at this filling is rather unexpected. 
Pan {\sl et al.} take their observations as evidence for residual 
CF--CF interactions, sufficiently strong to cause emergence of new 
FQH states (that, unlike virtually all observed fractions, cannot 
be predicted from the noninteracting CF model or standard Haldane 
hierarchy).
However, they ignore the theoretical investigations in which these 
interactions were studied in detail \cite{hierarchy,sitko,lee}.  

In this paper we propose an explanation for these new states 
involving formation of QP pairs which display Laughlin correlations 
with one another.
First, we explain the connection between the CF model and the QP 
hierarchy \cite{haldane,hierarchy,sitko}.
Second, we recall two simple types of two-body correlations,
Laughlin correlations and pairing, that may occur in an interacting 
system depending on the filling factor $\nu$ and on whether 
$V(\mathcal{R})$ is super- or subharmonic at the relevant range 
\cite{parentage}.
Then, knowing the QP--QP pseudopotential $V_{\rm QP}(\mathcal{R})$, 
we apply the concept of Laughlin condensed states of (bosonic) pairs
(used earlier for the electrons in the $n=1$ LL to describe such FQH 
states as $\nu=5/2$ or 7/3 \cite{fivehalf}) to the particles or holes 
in a partially filled CF LL, i.e., to Laughlin quasielectrons (QE's) 
or quasiholes (QH's).
Finally, we propose the novel hierarchy of FQH states in which 
the incompressibility results from the condensation of QP pairs 
(QE$_2$'s or QH$_2$'s) into Laughlin correlated pair states.
The series of FQH states derived from the parent $\nu=1/3$ state 
include all novel fractions: $\nu=5/13$, 3/8, 4/11, and 6/17 for 
the QE's and $\nu=5/17$, 3/10, 4/13, and 6/19 for the QH's.

\section{Laughlin QP's and CF's}

Let us begin with recalling the connection between Laughlin 
QP's \cite{laughlin} and the CF model (equivalent to the 
mean-field Chern--Simons transformation) \cite{jain}.
Laughlin QP's are the actual elementary excitations of
a two-dimensional electron liquid filling a fraction of
the lowest LL.
They have well-defined and known wave functions (and thus 
also density profiles, size, etc.) and single-particle energy.
They carry (fractional) electric charge, counted with 
respect to the uniformly spread charge of the underlying 
Laughlin state.
The negatively and positively charged QP's are called 
quasielectrons (QE's) and quasiholes (QH's), respectively.
Moving in the external magnetic field, both kinds of QP's 
follow cyclotron orbits, conveniently labeled by angular 
momentum.

Since the uniform-density Laughlin states only occur at a 
discrete series of filling factors ($\nu=1/3$, $1/5$, \dots),
and Laughlin QP's are simply the least-energy excitations 
able to carry charge in excess of the value corresponding 
to the nearest Laughlin state, their type and number 
depends on $\nu$.
At $\nu$ precisely equal to any of the Laughlin values 
$\nu_p=(2p+1)^{-1}$ no QP's exist in the ground state
and they may only appear in form of excitonic, 
charge-neutral QE--QH pairs.
At $\nu$ slightly smaller (or larger) than $\nu_p$, 
a number $N_{\rm QP}$ of the ``$p$--type'' QE's (or QH's) 
appear in the Laughlin liquid. 
The QP number is equal to the difference between the total 
magnetic flux through the sample and its value corresponding 
to the nearest ``parent'' Laughlin state, measured in units 
of elementary flux, $N_{\rm QP}=|\Phi-\Phi_p|/\phi_0$. 

For electrons, the filling factor is expressed through the 
flux $\Phi$ and the electron number $N$ as $\nu=\Phi/N\phi_0$,
For the QP's of the parent $\nu_p$ Laughlin state, each 
carrying charge $\pm e/(2p+1)$, the LL degeneracy is reduced 
by a factor of $(2p+1)$ compared to the electron value.
Consequently, the QP filling factor is equal to $\nu_{\rm QP}
=\Phi/(2p+1)N_{\rm QP}\phi_0$.
Note that $\nu_{\rm QP}$ is a linear function of $\nu$ near 
each value of $\nu_p$, 
The situation with $\nu$ far away from the nearest $\nu_p$, 
corresponds to a large filling factor for appropriate QP's 
whose mutual interactions and filling of higher QP LL's both 
become important.

In the mean-field CF picture, the reduction of the QP LL 
degeneracy is attributed to a reduced magnetic field 
$B^*=B/(2p+1)$ rather than to a reduced QP charge.
In the simplest formulation of the model, the effective field 
$B^*$ results from the capture of an even number of magnetic 
flux quanta $2p\phi_0$ by each electron (such bound state 
is called a CF), and the QE's and QH's are pictured as particles 
and vacancies in the otherwise empty (full) CF LL's.

The CF picture turns out very useful for the description
of many properties of the QP's (e.g., size of cyclotron
orbits or LL degeneracy) or of the FQH systems in general
(e.g., the values of $\nu$ at which the incompressibility
results from a complete filling of the QP/CF LL's).
However, the QP--QP interactions important for the present
problem in the CF model arise as a combination of rather 
complicated two- and three-body gauge interactions between 
charges and fluxes, and because of this difficulty, are 
usually neglected.
It is therefore important to realize that the QP--QP 
interaction really is a Coulomb interaction between a pair
of charged particles.
For example, for two identical QP's it is repulsive, and at 
long range it is similar to the electron--electron repulsion, 
only reduced in magnitude by $(2p+1)^2$ because of a smaller 
QP charge,
On the other hand, the exact form of the QP--QP interaction at 
short range (where it is different from the electron--electron 
repulsion because of the particular QP charge-density profile)
has been quite accurately calculated numerically \cite{hierarchy}.

\section{Haldane Hierarchy and Jain Sequence}

Knowing that the QE--QE and QH--QH interactions are generally
repulsive, Haldane proposed \cite{haldane} condensation of QE's 
and QH's into the hierarchy of ``daughter'' states at the series 
of Laughlin values of $\nu_{\rm QE}$ or $\nu_{\rm QH}$.
In these states, the appropriate QP's correlate with one another
in the same way as the electrons do in the Laughlin states, and
their elementary excitations are simply a new generation of QP's.
Assumming such QP condensation at each level of the hierarchy
one would predict incompressibility of the whole electron system
at all filling factors given by any odd-denominator fraction.
This is in striking disagreement with the experiments, and the
reason for this discrepancy is that although Coulombic, the 
QP--QP interaction at short range is not quite identical to 
the electron--electron repulsion responsible for the Laughlin 
correlations.
As a result, the QP's form Laughlin liquids only at very few of 
the Laughlin fractions, which eliminates all but a few valid 
``hierarchy'' fractions \cite{hierarchy}, in good agreement 
with the experiment.
The same series of fractions arise naturally in the CF picture.
These are the states at $\nu=(2p+1/n)^{-1}$, corresponding to 
a number $n$ of completely filled CF LL's each carrying flux 
$2p\phi_0$.

The new FQH states \cite{pan} occur at the values of $\nu$ from 
outside the Jain sequence, and thus corresponding to only partial 
filling of a CF LL.
Hence, in the CF picture, their incompressibility implies role
of CF--CF interactions.
In the QP hierarchy picture, the new states either coincide with 
the ``invalid'' fractions (e.g., $\nu=4/11$) or are new fractions 
altogether (e.g., $\nu=3/8$).
In both cases it is clear that the origin of observed 
incompressibility lies in the special form of QP--QP correlations, 
and that these correlations are of a new (non-Laughlin) type.

\section{QP--QP Pseudopotential}

The nature of QP correlations depends critically on the form 
of pseudopotential $V_{\rm QP}(\mathcal{R})$ describing their 
pair interaction energy $V_{\rm QP}$ as a function of relative 
pair angular momentum $\mathcal{R}$.
We have shown earlier \cite{parentage,fivehalf} that the 
correlations are of the Laughlin type (i.e., the particles tend 
to avoid pair states with one or more of the smallest values of 
$\mathcal{R}=1$, 3, \dots) only if $V(\mathcal{R})$ is 
``superharmonic'' at the relevant values of $\mathcal{R}$ for 
a given filling factor $\nu$ (specifically, at $\mathcal{R}=2p-1$ 
for $\nu\sim(2p+1)^{-1}$, where $p=1$, 2, \dots).
Laughlin correlations defined in this way justify reapplication 
of the CF picture to the QP's to select the lowest states of the 
whole many-body spectrum, and lead to the incompressible QP 
``daughter'' states of the standard CF hierarchy \cite{sitko}.
The superharmonic repulsion is defined as one for which $V$ 
decreases more quickly than linearly as a function of the average 
particle--particle separation $\left<r^2\right>$ for the consecutive 
pair eigenstates labeled by $\mathcal{R}$.
In spherical geometry \cite{haldane}, most convenient for finite-size 
calculations, this means that $V$ increases more quickly than linearly 
as a function of $L(L+1)$, i.e., of the squared total pair angular 
momentum $L=2l-\mathcal{R}$, where $l$ is the single-particle angular 
momentum.

The qualitative behavior of the QP--QP interaction pseudopotential 
$V_{\rm QP}(\mathcal{R})$ at short range is well-known from 
numerical studies of small systems \cite{hierarchy,sitko,lee}.  
On the other hand, the repulsive character of the QP--QP interaction
and the long-range behavior of $V_{\rm QP}(\mathcal{R})\sim
\mathcal{R}^{-1/2}$ follow from the fact that QP's are charged 
particles (the form of QP charge density affects $V_{\rm QP}$ 
only at short range, comparable to the QP size).
Combining the above arguments, it is clear that the dominant 
features of $V_{\rm QE}$ are the small value at $\mathcal{R}=1$ 
and a strong maximum at $\mathcal{R}=3$.
Analogous analysis for the QH's yields maxima at $\mathcal{R}=1$ 
and 5, and nearly vanishing $V_{\rm QH}(3)$.

\section{QP Pairing}

It is evident that $V_{\rm QE}$ does not support Laughlin QE--QE 
correlations.  
Instead, we expect that at least some of the QE's will form pairs 
(QE$_2$) at $\mathcal{R}=1$.
A paired state would be characterized by a greatly reduced fractional 
parentage $\mathcal{G}$ \cite{parentage} from the strongly repulsive 
$\mathcal{R}=3$ state compared to the Laughlin correlated state, and 
have lower total interaction energy $E={1\over2}N(N-1)\sum_\mathcal{R}
\mathcal{G}(\mathcal{R})V(\mathcal{R})$.
Let us stress that such pairing is not a result of some attractive
QE--QE interaction, but due to a tendency to avoid the most strongly 
repulsive $\mathcal{R}=3$ pair state.
At sufficiently high QE density this can only be achieved by having 
significant $\mathcal{G}(1)$, which can be interpreted as pairing 
into the QE$_2$ molecules.
By analogy, the QH pairing is expected at $\mathcal{R}=3$.
The range of $\nu_{\rm QP}$ at which pairing can be considered 
is limited by the condition that the separation between the 
pairs must exceed the pair size.
While for the QE pairs this is satisfied at any $\nu_{\rm QE}<1$, 
the QH pairing can only occur at $\nu_{\rm QH}<1/3$.

\section{Laughlin Correlations Between Pairs}

Having established that the QP fluid consists of (bosonic) 
QP$_2$ molecules, the QP$_2$--QP$_2$ interactions need be 
studied to understand correlations.
The QP$_2$--QP$_2$ interaction is described by an effective 
pseudopotential $V_{{\rm QP}_2}(\mathcal{R})$ that includes 
correlation effects caused by the fact that the two-pair 
wavefunction must be symmetric under exchange of the whole 
QP$_2$ bosons and at the same time antisymmetric under exchange 
of any pair of the QP fermions.
This problem is analogous to that of interaction between the 
electron pairs in the $n=1$ LL \cite{fivehalf}.

Although we do not know $V_{{\rm QP}_2}(\mathcal{R})$ accurately,
we expect that since it is due to the repulsion between the QP's 
that belong to different QP$_2$ pairs, it might be superharmonic 
at the range corresponding to the QP$_2$--QP$_2$ separation.
Our preliminary numerical results for four QE's seem to 
support this idea.
However, in contrast to the $n=1$ electron LL \cite{fivehalf},
the lack of accurate data for $V_{\rm QP}$ at the intermediate 
range makes such calculations uncertain.

\section{Condensed Pair States}

The assumption of Laughlin correlations between the QP$_2$ bosons 
implies the sequence of Laughlin condensed QP$_2$ states that can 
be conveniently described using the ``composite boson'' (CB) model 
\cite{fivehalf}.
Let us use spherical geometry and consider the system of $N_1$ 
fermions (QP's) each with (integral or half-integral) angular 
momentum $l_1$ (i.e., in a LL of degeneracy $g_1=2l_1+1$).
Neglecting the finite-size corrections, this corresponds to the 
filling factor $\nu_1=N_1/g_1$.
Let the fermions form $N_2={1\over2}N_1$ bosonic pairs each with 
angular momentum $l_2=2l_1-\mathcal{R}_1$, where $\mathcal{R}_1$ 
is an odd integer.
The filling factor for the system of pairs, defined as $\nu_2=N_2/g_2$ 
where $g_2=2l_2+1$, equals to $\nu_2={1\over4}\nu_1$.
The allowed states of two bosonic pairs are labeled by total 
angular momentum $L=2l_2-\mathcal{R}_2$, where $\mathcal{R}_2$ is an
even integer.
Of all even values of $\mathcal{R}_2$, the lowest few are not allowed
because of the Pauli exclusion principle applied to the individual 
fermions.
The condition that the two-fermion states with relative angular 
momentum smaller than $\mathcal{R}_1$ are forbidden is equivalent 
to the elimination of the states with $\mathcal{R}_2\le4\mathcal{R}_1$ 
from the two-boson Hilbert space.
Such a ``hard core'' can be accounted for by a CB transformation
with $4\mathcal{R}_1$ flux quanta attached to each boson\cite{theorem}.
This gives effective CB angular momentum $l_2^*=l_2-2\mathcal{R}_1
(N_2-1)$, LL degeneracy $g_2^*=g_2-4\mathcal{R}_1(N_2-1)$, 
and filling factor $\nu_2^*=(\nu_2^{-1}-4\mathcal{R}_1)^{-1}$.

The CB's defined in this way condense into their only allowed
$l_2^*=0$ state ($\nu_2^*=\infty$) when the corresponding fermion 
system has the maximum density at which pairing is still possible,
$\nu_1=\mathcal{R}_1^{-1}$.
At lower filling factors, the CB LL is degenerate and the spectrum
of all allowed states of the $N_2$ CB's represents the spectrum of 
the corresponding paired fermion system.
In particular, using the assumption of the superharmonic form of 
boson--boson repulsion, condensed CB states are expected at 
a series of Laughlin filling factors $\nu_2^*=(2q)^{-1}$.
Here, $2q$ is an even integer corresponding to the number of
additional magnetic flux quanta attached to each CB in a subsequent
CB transformation, $l_2^*\rightarrow l_2^{**}=l_2^*-q(N_2-1)$,
to describe Laughlin correlations between the original CB's of
angular momentum $l_2^*$.
From the relation between the fermion and CB filling factors,
$\nu_1^{-1}=(4\nu_2^*)^{-1}+\mathcal{R}_1$, we find the following 
sequence of fractions corresponding to the Laughlin condensed pair 
states, $\nu_1^{-1}=q/2+\mathcal{R}_1$.
Finally, we set  $\mathcal{R}_1=1$ for the QE's and $\mathcal{R}_1=3$ 
for the QH's, and use the hierarchy equation \cite{hierarchy}, 
$\nu^{-1}=2p+(1\pm\nu_{\rm QP})^{-1}$, to calculate the following 
sequences of electron filling factors, $\nu$, derived from the parent 
$\nu=(2p+1)^{-1}$ state
\begin{equation}             
   \nu^{-1}=2p+1\mp(2+q/2)^{-1},
\end{equation}
where ``$+$'' corresponds to the QE's and ``$-$'' to the QH's.
Remarkably, all the fractions reported by Pan {\sl et al.} are
among those predicted for the $\nu=1/3$ parent. 
Note also that the same values of $q=1$, 2, 4, and 8 describe
both observed QE and QH states.
This indicates similarity of the QE--QE and QH--QH pseudopotentials 
and suggests that both $V_{{\rm QE}_2}$ and $V_{{\rm QH}_2}$ may be
superharmonic only at the corresponding four values of $\mathcal{R}$
(in such case, the remaining fractions could not be observed even 
in most ideal samples).

\section{Conclusion}

We have studied the QP--QP interactions leading to novel 
spin-polarized FQH states in the lowest LL.
Using the knowledge of QP--QP pseudopotentials and a general 
dependence of the form of correlations on the super- or 
subharmonic behavior of the pseudopotential, we have shown 
that QP's form pairs over a certain range of filling factor 
$\nu_{\rm QP}$.
Then, we argued that the correlations between the QP pairs 
should be of Laughlin type and proposed a hierarchy of 
condensed paired QP states.
The proposed hierarchy of fractions agrees remarkably well 
with the recent experiment of Pan {\sl et al.} \cite{pan}.

\acknowledgments

The authors thank Jennifer J. Quinn and Josef Tobiska for helpful 
discussions and acknowlege support from US DoE grant Grant DE-FG 
02-97ER45657.
AW acknowledges support from KBN grant 2P03B02424.
KSY acknowledges support from KOSEF grant R14-2002-029-01002-0.

\end{document}